\documentclass{ws-p8-50x6-00}

\begin{document}

\title{Imaging Atmospheric \v{C}erenkov Telescopes: 
               Techniques and Results}

\author{S. M. Bradbury}

\address{Dept. of Physics \& Astronomy, University of Leeds, Leeds LS2 9JT,  
U.K.\\ 
E-mail: smb@ast.leeds.ac.uk}

\maketitle

\abstracts{ The hunt for cosmic TeV particle accelerators is prospering
        through Imaging Atmospheric \v{C}erenkov Telescopes. We face
        challenges such as low light levels and MHz trigger rates, and the
        need to distinguish between particle air showers stemming from primary
        $\gamma$-rays and those due to the hadronic cosmic ray background. Our
        test beam is provided by the Crab Nebula, a steady accelerator of
        particles to energies beyond 20\,TeV. Highly variable $\gamma$-ray
        emission, coincident with flares at longer wavelengths, is revealing
        the particle acceleration mechanisms at work in the relativistic jets
        of Active Galaxies. These 200\,GeV to 20\,TeV photons propagating over
        cosmological distances allow us to place a limit on the infra-red
        background linked to galaxy formation and, some speculate, to the
        decay of massive relic neutrinos.  $\gamma$-rays produced in
        neutralino annihilation or the evaporation of primordial black holes
        may also be detectable. These phenomena and a zoo of astrophysical
        objects will be the targets of the next generation multi-national
        telescope facilities.  }

\section{Introduction}

$\gamma$-ray astronomy has traditionally been thought of as an extension of
X-ray satellite experiments into the $\gamma$-ray regime. An all-sky survey by
the most recent satellite experiment, EGRET (1991 - 2000), increased the
catalogue of discrete astronomical sources of 20\,MeV - 20\,GeV $\gamma$-rays
from 4 to 271. As a result, the question of what happens at energies beyond
20\,GeV has gained momentum. Since the $\gamma$-ray count rate from these
sources is a steeply falling function of photon energy, a sensitive area four
orders of magnitude greater than EGRET's is needed for reasonable counting
statistics in the Very High Energy (200\,GeV - 50\,TeV) $\gamma$-ray
regime. To attain this we must make the absorbing atmosphere work for us
as our calorimeter.

In the early 1960s, it was suggested that cosmic photons at energies of
$\sim$1\,TeV from point sources could be detected via \v{C}erenkov radiation,
produced by the e$^-$ e$^+$ particle airshowers which they initiated in the
atmosphere. The main drawback of this method for the study of $\gamma$-rays
was the huge background of airshowers produced by charged cosmic rays.  It was
essential to be able to discriminate between the two classes of
primary particles.  Success came in 1989 with the introduction of the
\v{C}erenkov imaging technique, by which the Crab Nebula, the remnant of the
supernova of AD\,1054, was firmly established as a source of TeV
$\gamma$-rays.\cite{weekes89} Current atmospheric \v{C}erenkov detectors
operate either by imaging, as described here, or by wavefront
sampling.\cite{ong98}

Almost a century after their discovery, the origin of hadronic cosmic rays
remains a mystery. Their arrival directions at Earth provide no clue as to
their source since their paths are contorted by galactic magnetic fields; they
form an apparently isotropic background for \v{C}erenkov telescopes. The
hadronic cosmic ray distribution throughout the galaxy should be
traceable through their interactions with matter and subsequent $\gamma$-ray
emission, for example via $\pi^\circ$ decay. The remnants of supernova
explosions are generally believed to supply the cosmic ray particles of
energies up to $Z \times 10^{14}$\,eV in our galaxy, since little else could
provide sufficient energy. Such objects are therefore prime targets for VHE
$\gamma$-ray astronomers seeking the cosmic ray accelerators.

 Less than 1\% of the VHE $\gamma$-ray sky has been mapped by the most
sensitive means and there are about a dozen recognised VHE $\gamma$-ray
sources. Some are active galaxies, which emit dramatic flares during which the
count rate of VHE $\gamma$-rays doubles on a timescale of
hours.~\cite{gaidos}~\cite{quinn99} The $\gamma$-ray luminosity and flare
timescale, together with near simultaneous X-ray and optical emission
episodes, allow us to estimate the magnetic field strength and Doppler factor
at the source, indicating emission from highly relativistic plasma jets.

\section{The Imaging Atmospheric \v{C}erenkov Technique}

Typically about 0.01\% of the energy of an incoming cosmic $\gamma$-ray is
expected to be dissipated as \v{C}erenkov light. This illuminates a disc of
radius $\sim$125\,m on the ground for the $\sim$10\,ns equivalent to the
lifetime of the particle shower. In essence, an atmospheric \v{C}erenkov
telescope acts as a ``light bucket'', a single mirrored dish which reflects a
fraction of the \v{C}erenkov light pool onto a camera in the focal plane.

\subsection{Telescope Construction}

\begin{figure}[t]
\epsfxsize=20pc
\begin{center}
\epsfbox{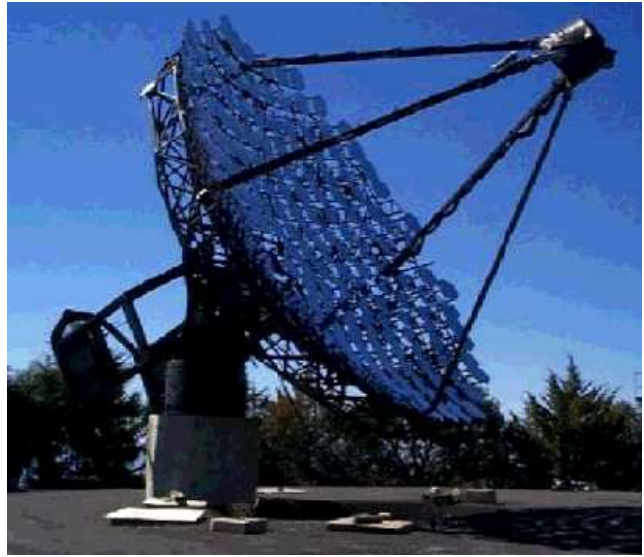}
\end{center}
\caption{The Whipple Observatory 10\,m imaging atmospheric \v{C}erenkov telescope.\label{fig:whipplet}}
\end{figure}

The largest single dish now in operation is the 10\,m diameter telescope at
the Whipple Observatory in Arizona, shown in figure \ref{fig:whipplet}.  This
will be surpassed by the 17\,m diameter MAGIC Telescope in the latter half of
2001.  Atmospheric ozone absorbs much of the UV component of the \v{C}erenkov
flash so that the light intensity peaks at around 300\,nm, for which a high
reflectivity can be achieved with an anodized aluminium coating on a glass or
lightweight aluminium substrate.  Current telescopes use cameras of several
hundred close-packed photomultiplier tubes to record images of the
\v{C}erenkov light flashes. These are digitized and subsequently parameterized
in terms of brightness, shape, orientation and angular position. The ability
to derive an arrival direction, primary energy and primary particle type
($\gamma$-ray, cosmic ray nucleon or local muon) from these images is
dependent upon both the accuracy of our models of high energy particle cascade
development, and our knowledge of the variable state of the calorimetric
volume of atmosphere above the telescope.

\subsection{Nature's Challenges}

\begin{figure}[t]
\epsfxsize=20pc
\begin{center}
\epsfbox{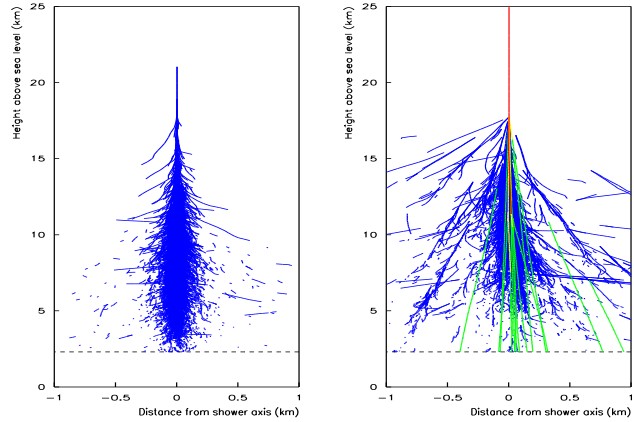}
\end{center}
\caption{Simulated air showers initiated by vertical 1\,TeV $\gamma$-ray 
(left) and proton (right) primaries. Few particles reach the altitude 
of the Whipple Observatory (dashed line). \label{fig:showers}}
\end{figure}

\begin{figure}[t]
\epsfxsize=25pc
\begin{center}
\epsfbox{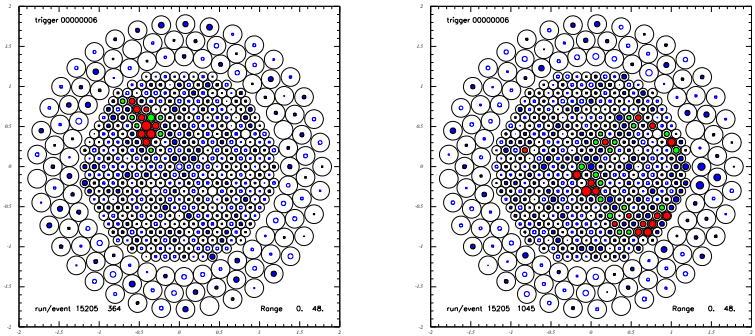}
\end{center}
\caption{\v{C}erenkov events recorded by the Whipple Telescope attributed
to a $\gamma$-ray (left) and a single local muon (right). Filled circles 
scale with no. of ADC counts in a channel. \label{fig:events}}
\end{figure}

\v{C}erenkov telescopes are pointed instruments viewing a few square degrees
of sky. $\gamma$-rays from an astronomical object are expected to produce
compact \v{C}erenkov images pointing towards the position of that object in
the field of view. Cosmic ray induced airshowers develop sub-showers as a
result of the relatively high transverse momentum of hadronic daughter
particles and should therefore produce identifiably irregular \v{C}erenkov
images. When attempting to lower a telescope's energy threshold, by reducing
the threshold light level for an event to be recorded, individual muons
created in hadronic showers which reach the level of the detector (as seen in
figure \ref{fig:showers}) can mimic $\gamma$-ray events. Whilst the
\v{C}erenkov light ring from a muon striking the centre of the mirror is
clearly recognisable (in figure \ref{fig:events}), a small ``muon arc'' image
will be recorded if the impact point is several metres from the telescope. One
way to eradicate this background is to require that a $\gamma$-ray like image
has been simultaneously observed by multiple telescopes tracking the same
astronomical object.~\cite{daum} A further source of background noise events
is quite simply the fluctuating brightness of the night sky. This can
instantaneously result in two or three random photomultiplier signals above
the fixed discrimination threshold, triggering the recording of a false
event. A programmable hardware trigger has been developed for the Whipple
Telescope which can identify candidate \v{C}erenkov events with {\em adjacent}
pixel signals within 65\,ns.~\cite{bradbury2001}

Dealing with background light is further complicated by bright stars in the
field of view. As objects are tracked across the sky the movement of the
telescope places a variable strain on cables and connectors. These must also
withstand diurnal and seasonal temperature changes, humidity, wind and an
occasional loading of snow at high altitudes. Where necessary control
connections are made over optical fibre to reduce the likelihood of lightning
damage.  Prototype, low cost, analogue fibre optic links have been developed
to transmit photomultiplier signals to data acquisition electronics on the
ground.~\cite{rose} These introduce far less signal dispersion than co-axial
cable. Keeping the pulse width narrow allows one to reduce the integration
time and hence the amount of night sky background noise included with the
\v{C}erenkov signal.~\cite{fibres} The use of FADCs for charge digitization
can provide not only a better estimate of instantaneous background light
levels, but also some information on the time structure of each pulse, a
characteristic of the airshower's development and hence a potential clue to
the primary particle type.

\section{Cosmic Particle Accelerators}

\subsection{Galactic VHE $\gamma$-ray Sources}

\begin{figure}[t]
\epsfxsize=25pc
\begin{center}
\epsfbox{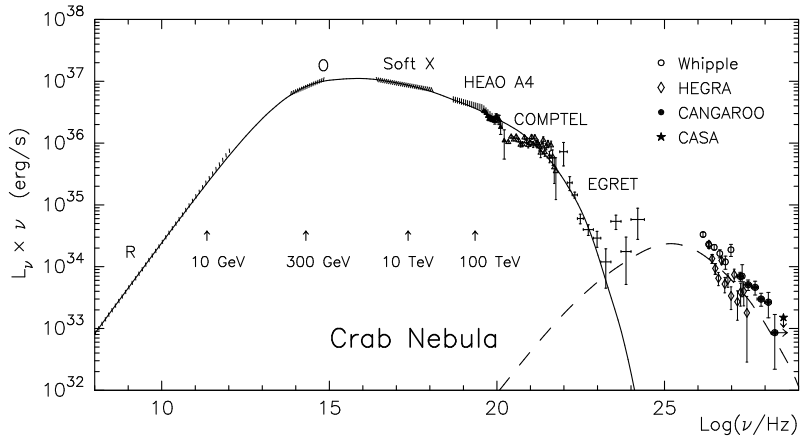}
\end{center}
\caption{Nonthermal radiation of the Crab Nebula. The solid
and dashed curves are model fits of synchrotron 
and inverse Compton components by Aharonian and Atoyan$^1$.\label{fig:crab}}
\end{figure}

The non-thermal radiation from the Crab Nebula supernova remnant is well
documented from radio wavelengths through to VHE $\gamma$-rays.  It is
supposed that the central neutron star spinning at 30\,Hz is generating a
pulsar wind of relativistic electrons.  The spectrum, shown in figure
\ref{fig:crab} is dominated by the interactions of these 
electrons with the magnetic fields in the gaseous nebula and ambient photon
fields.  Synchrotron emission can account for the curve from the radio through
to satellite $\gamma$-ray observations by COMPTEL and EGRET at up to
1\,GeV. The corresponding electron energies required are from 10\,GeV to
100\,TeV as indicated.  The VHE $\gamma$-ray emission is fitted by a second
component (dashed curve in figure \ref{fig:crab}), due to the inverse Compton
scattering of soft photons up to energies above 10\,TeV by relativistic
electrons.~\cite{crab} Effective acceleration of particles to beyond
10$^{14}$\,eV is implicit.  Given the electron spectrum required to generate
the VHE $\gamma$-ray flux, a local magnetic field strength of $\sim$16\,nT was
deduced from the observed level of X-ray synchrotron emission presumed to be
due to these same electrons.~\cite{hillas98}

There have been three tentative detections of VHE $\gamma$-rays from
shell-type supernova remnants which lack a central neutron star dynamo.  X-ray
observations of one of these objects, SN\,1006, indicate synchrotron emission
from electrons with energies of $\sim$100\,TeV. The $\gamma$-ray flux can
therefore be accounted for without invoking $\gamma$-ray production as a
result of collisions between relativistic protons and the interstellar medium
(principally atomic hydrogen) e.g. according to $p + ISM \rightarrow \pi^\circ
+ X \rightarrow \gamma + \gamma + X $. There is still no firm evidence for
hadronic cosmic ray production at supernova shock fronts.

\subsection{$\gamma$-ray Sources at Cosmological Distances}

There are two well-established sources of VHE $\gamma$-rays outside our own
galaxy, objects known as Markarian\,421 and Markarian\,501. These are both
blazar type active galaxies, at a distance of some 500 million light years, in
which gravitational accretion of matter onto a central supermassive black hole
is assumed to be powering an outflow of material in relativistic plasma jets.
\cite{agnpic} The rapid variability and high luminosity of the VHE
$\gamma$-ray flux from these objects seems to indicate that the $ \gamma$-rays
are relativistically beamed. Our line of sight to each must therefore be such
that we happen to be looking straight down one of the jets. For the 15 minute
timescale VHE $\gamma$-ray variability observed from Markarian\,421, causality
implies an emission region as small as our solar system.~\cite{gaidos}

The multi-wavelength spectra of Markarian\,421 and Markarian\,501 conform
quite well to the two component synchrotron plus electron inverse Compton
scattering mechanism proposed for the Crab Nebula. Correlated episodes of
enhanced emission of $\gamma$-ray and optical/UV photons imply that these
photons are produced in the same region of the jet. In this case, jet Doppler
factors in the range of 2 $< \delta >$ 40 are required to avoid a significant
loss of VHE $\gamma$-rays to pair production with the low energy photons. If
the VHE $\gamma$-rays in fact result from hadronic cascades produced by high
energy protons in the jet, as proposed by Mannheim~\cite{mannheim} and others,
then $\delta \approx 10$.
 
 The cosmic infrared background (CIB) traces the history of star formation and
galaxy evolution in the early universe. VHE $\gamma$-rays from active
galaxies can provide a means of probing the CIB. The $\gamma$-ray signal will
be attenuated by pair production: \mbox{$\gamma_{VHE} + \gamma_{IR}
\rightarrow e^- + e^+ $.} Since the attenuation becomes more significant as
the $\gamma$-ray energy increases, we expect to see a distance dependent, high
energy cut-off in the VHE spectra of active galaxies. Upper limits on the CIB
which are more restrictive than direct measurements in the 0.025 to 0.3\,eV
range have already been obtained.~\cite{irb} To establish whether spectral
cut-offs are due to this effect or are in fact intrinsic to these objects more
sensitive instruments are required to extend the source catalogue.

\section{Next Generation Telescopes}

Several ``next generation'' atmospheric \v{C}erenkov observatories are
under construction. VERITAS~\cite{weekes} in Arizona, HESS in Namibia and
CANGAROO III in Australia will be arrays of 10\,m diameter class imaging
\v{C}erenkov telescopes.  By viewing each airshower with several telescopes
separated by about 100\,m, greatly improved angular and energy resolution and
an increased collection area will be achieved between 100\,GeV and 10\,TeV.
As a single 17\,m diameter dish, the MAGIC Telescope, on the Canary
Island La Palma, will have the lowest energy threshold of 30\,GeV using a
standard photomultiplier camera, or 15\,GeV if equipped with hybrid
GaAsP photocathode photodetectors.~\cite{barrio}

\subsection{Fundamental Physics}

The greater sensitivity of the planned instruments will improve the chances of
new detections of active galaxies, the pulsed signatures of objects such as
the Crab pulsar and $\gamma$-ray bursts. Their angular resolution should
enable us to identify some of the 170 sources detected with large position
errors by the EGRET satellite instrument. In addition, VHE $\gamma$-ray
observations may improve our view of some fundamental physics
phenomena.

Our current understanding requires the presence of cold dark matter in the
universe to explain certain astrophysical data. The neutralino is a favoured
dark matter candidate. A concentration of neutralinos towards the centre of
our galaxy may produce a detectable monoenergetic annihilation line in the
neutralino mass range of 30\,GeV to 3\,TeV.~\cite{bergstrom}

$\gamma$-ray observations of distant objects can be searched for the effects
of quantum gravity. The velocity of light may exhibit an energy dependence due
to quantum fluctuations in a gravitational medium, resulting in a time
dispersion within VHE $\gamma$-ray flares. A 15 minute flare from Markarian
421 has already been used to place a limit on this effect,~\cite{biller99}
which could be vastly improved by instruments sensitive to minute by minute
variability.

Low mass black holes, remnants of inhomogeneities in the early universe,
should emit a burst of radiation peaked at around 1\,TeV in the final stages
of their evaporation, according to Halzen et al.~\cite{halzen} In fact, the
predicted time profile and energy spectrum depend on whether one follows the
standard or bootstrap models for the particle spectrum at high energies.

These are just a few of the topics in particle physics and astrophysics which
the next generation of imaging \v{C}erenkov telescopes may address.  I would
like to thank all involved in this Symposium for their encouragement!

\end{document}